\documentstyle[aps,emlines]{revtex}
\begin{document}
\draft
\title{On the Darwin Lagrangian}
\author{E.G.Bessonov}
\address{Lebedev Physical Institute, RAS, Leninsky pr. 53, Moscow
117924, Russia}
\date{23 February 1999}
\maketitle

                        \begin{abstract}
In this paper we explore some surprising consequences of the
retardation effects of Maxwell's electrodynamics to a system of charged
particles. The specific cases of three interacting particles are
considered in the framework of classical electrodynamics. We show that
the solutions of the equations of motion defined by the Darvin
Lagrangian in some cases contradict to common sense.  \end{abstract}


Darvin Lagrangian for interacting particles is an approximate one first
derived by Darvin in 1920 \cite{darvin}. This Lagrangian is considered
to be correct to the order of $1/c^2$ inclusive \cite{landau},
\cite{jackson}, \cite{coleman}. To this order, we can eliminate the
radiation modes from the theory and describe the interaction of charged
particles in pure action-at-a-distance terms. Although the Darwin
Lagrangian has had its most celebrated application in the
quantum-mechanical context of the Breit interaction, it has uses in the
purely classical domain \cite{jackson} - \cite{de-luca}. In this paper
we explore some surprising consequences of the retardation effects of
Maxwell's electrodynamics to a system of charged particles. The
specific cases of three interacting particles are considered in the
framework of classical electrodynamics.

Below we will present the detailed and typical derivation of the
Darvin Lagrangian and Hamiltonian for a system of charged particles to
correct some misprints made by authors. Then we will show that the
solutions of the equations of motion defined by the Darvin Lagrangian
in some cases contradict to common sense.

The Lagrangian for a particle of a charge $e_a$ in the external field
of an another particle of a charge $e_b$ is

\begin{equation}     
L_a = - m_ac^2/\gamma _a - e_a \phi _b + {e_a\over c}\vec A_b
\cdot\vec v_a,
\end{equation}
where $m_a$ is the mass of the particle $a$, $c$ the light velocity,
$\gamma _a = 1/\sqrt{1 - \beta_a^2}$ relativistic factor of the particle
$a$, $\beta _a = |\vec v _a/c|$, $\vec v_a$ the vector of a velocity of
the particle $a$, $\phi _b$ and $\vec A_b$ the scalar and vector
retarded potentials produced by the particle $b$.

The scalar and vector potentials of the field produced by the charge
$b$ at the position of the charge $a$ can be expressed in terms of the
coordinates and velocities of the particle $b$ (for $\phi _b$ to the
terms of order $(v_b/c)^2$, and for $\vec A _b$, to terms $(v _b/c$)

\begin{equation}   
\phi _b = {e_b \over R_{ab}}, \hskip 2cm \vec A _b = {e _b[\vec v_b +
(\vec v_b\cdot \vec R _{ab})\vec R _{ab}/R _{ab}^2]\over 2cR _{ab}},
\end{equation}
where $R_{ab} = |\vec R_{ab}|$, $\vec R_{ab} = \vec R_a - \vec R_b$,
$\vec R_{a}$ and $\vec R_{b}$ are the radius-vectors of the particles
$a,b$ respectively, $v _b = |\vec v_b|$, $\vec v_b$ the vector of a
velocity of the particle $b$ \cite{landau}.

Substituting these expressions in (1), we obtain the Lagrangian $L_a$
for the particle $a$ (for a fixed motion of the other particles $b$).
The Lagrangian of the total system of particles is

             \begin{equation} 
      L = L^{p} + L^{int},
              \end{equation}
where the Lagrangian of the system of free particles $L^p$ and the
Lagrangian of the interaction of particles $L^{int}$ are

     $$L^p = -\sum _a m_ac^2/\gamma _a \simeq  -\sum _a m_ac^2 + \sum
       _a \frac{m_ac^2\beta ^2}{2} + \sum _a \frac{m_ac^2\beta ^4}{8},$$

      $$L^{int} = - \sum _{a>b} {e _ae _b\over R_{ab}} + \sum _{a>b}
         {e_a e_b\over 2R_{ab}} \vec \beta _a \vec \beta _b + \sum
         _{a>b} {e_a e_b\over 2R_{ab}^3}(\vec \beta _a \vec
         R_{ab})(\vec \beta _b \vec R_{ab}). $$

The equation of motion of a particle $a$ is described by the equation
${d\vec P_{a}/dt} = {\partial L}/{\partial \vec R_a}$,
where $\vec P_{\alpha} = {\partial L}/{\partial \vec v_a}$ is the
canonical momentum of the particle. This equation according to (3) can
be presented in the form (see Appendix A)

$$\frac{d\vec p_a}{dt} = \sum _{a>b} {e _ae _b\over R_{ab}^3} (1 - \vec
\beta _a \vec\beta _b )\vec R_{ab} + \sum _{a>b} {e_a e_b\over
R_{ab}^3} (\vec R_{ab} \vec \beta _a) \vec \beta _b + \sum _{a>b} {e_a
e_b\over 2R_{ab}^3} \beta _b^2 \vec R_{ab} $$

\begin{equation} 
- \sum _{a>b} {3e_a e_b\over 2R_{ab}^5} (\vec R_{ab} \vec \beta _b)^2
\vec R _{ab} - \sum _{a>b} {e_a e_b\over 2c} [{\dot {\vec \beta b} \over
R _{ab}} + {(\vec R _{ab}\dot {\vec \beta _b}) \vec R _{ab}\over R
_{ab}^3}].
\end{equation}
where $\vec p_a = m_a\gamma _a\vec v _a$ is the kinetic (non-canonical)
momentum of the particle $a$.

The Hamiltonian of a system of charges in the same approximation must
be done by the general rule for calculating $H$ from $L$ ($H = \vec
v_{a}\vec P_{a} - L$). According to (3) (see Appendix A) the value

             \begin{equation} 
             H = H^{p} + H^{int},
             \end{equation}
where

     $$H^p  = \sum _a m_ac^2\gamma _a = \sum _a \sqrt {m_a^2 c^4 +
     p_a^2 c^2} \simeq \sum _a m_ac^2 + \sum _a \frac{p_a^2}{2m_a} -
     \sum _a \frac{p_a^4}{8c^2m_a^3},$$

      $$H^{int} =  \sum _{a>b} {e _ae _b\over R_{ab}} + \sum _{a>b}
         {e_a e_b\over 2c^2m_am_bR_{ab}} \vec p _a \vec p _b + \sum
         _{a>b} {e_a e_b\over 2c^2m_am_bR_{ab}^3}(\vec p _a \vec
         R_{ab})(\vec p _b \vec R_{ab}).$$

The constant value $\sum _a m_ac^2$ in (5) can be omitted. Here we
would like to note that contrary to \cite{landau} the last two items in
the term $H^{int}$ of the equation (5) has the positive sign and the
momentum $\vec p_a = m_a\gamma _a\vec v_a$ includes $\gamma $-factor
of the particle ($\gamma \simeq 1 + \beta ^2/2 + 3\beta ^4/8$). The
Hamiltonian expressed through the canonical momentum has the form (5),
where the ordinary momentum $\vec p_a$ is replaced by the canonical one
$\vec P_a$  and the signs of the last two terms are changed
\cite{darvin}\footnote {In \cite{landau} the Hamiltonian includes
small letters for momentum $\vec p_a = m_a\vec v_a$ that is $\vec p_a$
in \cite{landau} is the kinetic momentum. It differ from (5) because of
its derivation is based on erroneous connection of small corrections to
Lagrangian and Hamiltonian. If the Lagrangian have the form $L = L_0 +
L _1$ then without any approximation $H = H _0 + H_1$, where $H _0 =
\sum _{a>b} \vec v _a \vec P _{a0} - L _0$, $H _1 = \sum _{a>b} \vec v
_a \vec P _{a1} - L _1$, $ \vec P _{a} =  \vec P _{a0} +  \vec P _{a1}
$, $\vec P _{a0} = \partial L _0/\partial \vec v _a$, $\vec P _{a1} =
\partial L _1/\partial \vec v _a$ is the extra term to the canonical
(conjugate) momentum. In \cite{landau} this connection was used but the
term $\sum _{a>b} \vec v _a \vec P _{a1}$ was omitted. In our case this
term differ from zero as $L_1$ depends on velocity. At the same time if
we will start from the definition $H = \vec v_a\partial L/\partial \vec
v_a - L$ then we will receive (5) \cite{darvin}.}. When the particles
are moving in the external electromagnetic field then the term $\sum _a
e_a \phi - e_a(\vec P_a\vec A)/m_a c + e_a^2 |\vec A|^2/2m_ac^2$ is
included in the Hamiltonian, where $\phi$ and $\vec A$ are the external
scalar and vector potentials. In \cite{darvin} the term $e_a^2 |\vec
A|^2/2m_ac^2$ is omitted.

The Lagrangian (3) does not depend on time. That is why the Hamiltonian
(5) is the energy of the system \cite{darvin}.

Further we consider a special case when particles are moving along the
axis $x$ (see Fig.1). In this case the Lagrangian and Hamiltonian of
the system of particles are described by the expressions

             \begin{equation} 
             L = -\sum _a m_ac^2/\gamma _a - \sum _{a>b} {e _ae _b\over
             R_{ab}}(1 - {\beta _a \beta _b\over 2}),
             \end{equation}

             \begin{equation} 
             H = \sum _a \sqrt {m_a^2 c^4 + p_a^2 c^2} + \sum _{a>b} {e
             _ae _b\over R_{ab}}(1 + {p _a p _b\over c^2m_am_b}),
             \end{equation}
where $\beta _i, p _i$ are the x-components of the particle relative
velocity and kinetic momentum respectively.

The x-component of the force applied to the particle $a$ from the
particle $b$ according to (4) in this case is $dp_a/dt = e_a e_b/R
_{ab} ^2\gamma _b^2 - e_ae_b \dot \beta_b/cR_{ab}$. This force
corresponds to the electric field strength $\vec E_b = - \nabla \phi _b
- (1/c)(\partial \vec A_b/\partial t)$ produced by the particle $b$ and
determined by the equations (2).

As was to be expected in the case of the uniform movement of the
particle $b$ ($\dot \beta_b = 0$; the case $m _b\gg m_a$) the
electric field strength produced by the particle $b$ in the direction
of its movement is $\gamma ^2_b$ times less then in the state of rest.

Next we consider the dynamics of three particles $a$, $b$, $d$
according to the Darvin Lagrangian and Hamiltonian. Let particles $a$,
$b$ have charges $e_a = e_b = e > 0$, masses $m_a = m_b = m$ and
velocities $v_a = - v_b = v = c\beta $.  The particle $d$ is located at
the position $x = 0$ at rest ($v_d = 0$) its charge and mass are $q,
M$.

In this case the Hamiltonian is the energy of the system which
according to (7) can be presented in the form

          \begin{equation} 
          H = Mc^2 + 2mc^2\gamma _0 = Mc^2 + 2mc^2\gamma + \frac
          {e^2}{2R\gamma ^2} + \frac {2eq}{R}, \end{equation}
where $\gamma _0$ is the initial relativistic factor of the particles
$a, b$ corresponding to the limit $R \to \infty$, $R = |\vec R_a|$ the
distance between the particle $a$ and the origin of the coordinate
system.

It follows from the equation (8) the dependence between the distance
 $R$ and the $\gamma $-factor of the particles $a,b$

         \begin{equation} 
         R =  \frac {e^2/2\gamma ^2 +
         2eq}{2mc^2(\gamma _0 - \gamma)}.  \end{equation}

1. We can see that when $q > -e/4\gamma _0^2$ then the turning
point exist at which $p = v =0$ and $\gamma = 1$. According to (8) the
minimal distance between particle $a$ and the origin of the coordinate
system

\begin{equation} 
R _{min} =  \frac {e^2 + 4eq}{4mc^2(\gamma _0 - 1)} =
\frac {e^2 + 4eq}{4T_0},
\end{equation}
where $T_0$ is the initial kinetic energy of the particle $a$. The
value $eU = (e^2 + 4eq)/2R_{min}$ is the potential energy of two
particles at rest at the position of the turning point. According to
(10) the value $R_{min} > r_a/2$, where $r_a = e^2/m_a c^2$ is the
classical radius of the particle $a$.

In that case according to (10) and in conformity with the energy
conservation law the potential energy of two particles at the turning 
point is equal to the 
initial kinetic energy of the particles $2T_0$. Retardation does not
lead to any results which are contradict to common sense. The term in
the electric field strength and in the force (4) which is determined by
the acceleration will compensate the decrease of the repulsive forces
corresponding to the uniformly moving particles.

\vspace{25mm}
\special{em:linewidth 0.4pt}
\unitlength 1.00mm
\linethickness{0.4pt}
\begin{picture}(140.00,70.00)
\put(140.00,80.00){\vector(1,0){0.2}}
\emline{20.00}{80.00}{1}{140.00}{80.00}{2}
\emline{80.00}{78.67}{3}{80.00}{81.33}{4}
\put(120.00,80.00){\oval(1.33,2.67)[]}
\put(40.00,80.00){\oval(1.33,2.67)[]}
\put(55.00,80.00){\vector(1,0){0.2}}
\emline{40.00}{80.00}{5}{55.00}{80.00}{6}
\put(105.00,80.00){\vector(-1,0){0.2}}
\emline{120.00}{80.00}{7}{105.00}{80.00}{8}
\put(136.67,75.00){\makebox(0,0)[cc]{$x$}}
\put(120.00,75.00){\makebox(0,0)[cc]{$a$}}
\put(105.00,75.00){\makebox(0,0)[cc]{$\vec v_a$}}
\put(80.00,75.00){\makebox(0,0)[cc]{$0$}}
\put(55.00,75.00){\makebox(0,0)[cc]{$\vec v_b$}}
\put(40.00,75.00){\makebox(0,0)[cc]{$b$}}
\put(80.33,65.00){\makebox(0,0)[cc]{Fig.1. A scheme of two
particle interaction.}}
\put(40.00,85.00){\makebox(0,0)[cc]{$e_b$}}
\put(120.00,85.00){\makebox(0,0)[cc]{$e_a$}}
\end{picture}
\vspace{-60mm}

2. When $q = -e/4\gamma _0^2$ then according to (4), (7) the particles
$a,b$ are moving uniformly ($\dot \beta = 0$, $v = v_0, \gamma = \gamma
_0$). In that case particles can reach the distance $R =x = 0$, which is
not reachable for them under the condition of the same energy expense
$2T_0$ and a non-relativistic bringing closer of the particles. This
conclusion is valid in the arbitrary relativistic case as in this case
there is no emission of the electromagnetic radiation. It contradicts
to common sense as the particles can be stopped at any position $R$ to
give back the kinetic energy $2T_0$ (in the form of heat and so on) and
moreover contrary to the energy conservation law they will produce an
extra energy $eU(R)$ under the process of slow moving aside of these
particles under conditions of repulsive forces.

3. When $-e/4 < q < -e/4\gamma _0^2$ then the particles $a,b$ will be
brought closer under the condition of an acceleration by attractive
forces and "fall in" toward each other. At the same time under such
value of charge $q$ of the particle $d$ in the non-relativistic case
the particles $a$, $b$ will repel each other such a way that the
position $R = x =0$ will not be reachable for them if the same energy
expense $2T_0$ will be used for slow bringing closer of the particles.
In that case we have the same result which contradicts to common sense
as well.

4. When $q = - e/4$, $\gamma _0> 1$ then the particles will acquire an
additional energy when bringing closer. After stop by extraneous
forces at any position $R$ to give back the kinetic energy $2T > 2T_0$
the particles will not experience any force.

5. When $q < -e/4$ then the particles will acquire the higher value of
the energy then necessary for non-relativistic separation of the
particles. The particles can be stopped by extraneous forces at some
distance between them and then separated. Some gain of energy will take
place as well.

In the cases (3), (5) the velocities of particles may be compared with
the light velocity when Darwin Lagrangian does not valid because of
in that case the radiation can not be neglected. But the process of
"fall in" will be kept. In the case (5) the unphysical solution can
appear when particles will be stopped at the distance $R \ll r_e$ and
the total energy of the system at this position (new Hamiltonian) will
be negative ($eU(R) + Mc^2 + 2m_ac^2 <0$). This result is the known
fact for a system of two particles of the opposite sign which is beyond
of the present consideration.

This curious results are the reminiscent of the non-consistency of the
classical Maxwell-Lorentz electrodynamics. The existence of these
solutions is a genuine effect of electrodynamics with retardation.

I acknowledge discussions with A.I.Lvov.

\appendix

\section{}

The canonical momentum of the particle $a$ is

             \begin{equation} 
             \vec P_a = {\partial{L}\over \partial {\vec v_a}} =
             \vec p _a + \Delta \vec p _a,
             \end{equation}
where
             $$\Delta \vec p_a = \sum _{b\ne a} {e_a e_b\over
2c}[{\vec \beta _b \over R_{ab}} + {\vec R_{ab}(\vec R_{ab} \vec \beta
_b)\over R_{ab}^3}]. $$

The time derivative of the canonical momentum is
             \begin{equation} 
            {d\vec P_{a}\over dt} = {d\over dt}{\partial{L}\over
\partial {\vec v_a}} = \dot {\vec p _a} + \Delta \vec F _a,
\end{equation}
where $\dot {\vec p _a} = d\vec p _a/dt$, $\Delta \vec F _a = d(\Delta
\vec p _a)/dt$ or

 $$\Delta \vec F _a = \sum _{b\ne a} {e_ae_b\over 2R_{ab}^3}[\vec
R_{ab}(\vec \beta _a - \vec \beta _b, \vec \beta _b) + (\vec \beta _a -
\vec \beta _b)(\vec R_{ab} \vec \beta _b)] - \vec \beta _b(\vec
R_{ab},\vec \beta _a - \vec \beta _b)]$$ $$- \sum _{b\ne
a}{3e_ae_b\over 2R_{ab}^5}(\vec R_{ab} \vec \beta _b)(\vec R_{ab},\vec
\beta _a - \vec \beta _b) \vec R_{ab} - \sum _{b\ne a}{e_ae_b\over 2c}
[{\dot {\vec \beta _b} \over R_{ab}} + {\vec R_{ab}(\vec R_{ab}
\dot{\vec \beta_{b}})\over R^3_{ab}}].$$

The directional derivative of the Lagrangian is

             \begin{equation} 
{\partial L\over \partial \vec R _a} = \sum _{a>b} {e_ae_b\vec
R_{ab}\over R_{ab}^3} (1 - {\vec \beta _a \vec \beta _b\over 2}) +
\sum _{a>b} {e_ae_b\over 2R_{ab}^3}[\vec \beta _a(\vec R_{ab} \vec
\beta _b) + \vec \beta _b(\vec R_{ab} \vec \beta _a)] - \sum _{a>b}
{3e_ae_b\over 2R_{ab}^5}\vec R_{ab}(\vec R_{ab} \vec \beta _a)(\vec
R_{ab} \vec \beta _b).  \end{equation}

From the equation of motion and equations (A2),(A3) it follows the
equation (4).

The value $\vec v_k \vec P_k$  and the Hamiltonian are equal
respectively

             \begin{equation} 
             \vec v _a \vec P _a = \sum _{a \ne b} \frac
{e_ae_b}{2}[\frac {\vec \beta _a \vec \beta _b}{\vec R_{ab}} +
\frac{(\vec \beta _a \vec R_{ab})(\vec \beta _b \vec R
_{ab})}{P_{ab}^3}] + m _a c^2 \gamma _a \beta _a^2,  \end{equation}

 $$ H = \sum _a \vec v_a \vec P _a - L = \sum _a \sqrt {m _a^2 c^4
+ p _a^2 c^2} +  \sum _{a>b} \frac {e_ae_b}{R _{ab}}[1 + \frac {c^2
(\vec p_a \vec p_b)}{2 \sqrt {m _a^ 2c^4 + c^2p_a^2} \sqrt{m _b^2 c^4 +
c^2p^2_b}} $$

             \begin{equation} 
+ \frac {c^2 (\vec R _{ab} \vec p_a)(\vec R_ {ab}\vec
p_b)}{2 R _{ab}^2\sqrt {m _a^ 2c^4 + c^2p_a^2}\sqrt {m _bc^4 +
c^2p^2_b}}]. \end{equation}

In the approximation ($1/c^2$) the Hamiltonian (A5) leads to (5).

\end{document}